\documentclass[conference]{IEEEtran}
\IEEEoverridecommandlockouts

\usepackage{cite}
\usepackage{amsmath,amssymb,amsfonts}
\usepackage{multirow}
\usepackage{booktabs}
\usepackage{algorithmic}
\usepackage{graphicx}
\usepackage{textcomp}
\usepackage{xcolor}
\def\BibTeX{{\rm B\kern-.05em{\sc i\kern-.025em b}\kern-.08em
    T\kern-.1667em\lower.7ex\hbox{E}\kern-.125emX}}
\begin{document}

\title{Decoupled Multimodal Fusion for User Interest Modeling in Click-Through Rate Prediction\thanks{\textsuperscript{*}Corresponding author.}}

\author{\IEEEauthorblockN{
Alin Fan\IEEEauthorrefmark{2},
Hanqing Li\IEEEauthorrefmark{2},
Sihan Lu\IEEEauthorrefmark{3},
Jingsong Yuan\IEEEauthorrefmark{2},
Jiandong Zhang\IEEEauthorrefmark{2}\textsuperscript{*}
}
\IEEEauthorblockA{\IEEEauthorrefmark{2} \textit{Alibaba International Digital Commerce Group}, Beijing, China}
\IEEEauthorblockA{\IEEEauthorrefmark{3} \textit{Renmin University of China}, Beijing, China}
\IEEEauthorblockA{
alin.fal@alibaba-inc.com,
lihanqing.lhq@alibaba-inc.com,
lusihan103278@ruc.edu.cn\\
jingsong.yjs@alibaba-inc.com,
chensong.zjd@taobao.com}
}

\maketitle

\begin{abstract}
Modern industrial recommendation systems enhance Click-Through Rate (CTR) prediction by integrating frozen multimodal representations from pre-trained models into ID-based frameworks. However, existing approaches predominantly adopt modality-centric modeling strategies that process ID and multimodal signals in isolation, thereby failing to capture fine-grained interactions between behavioral dynamics and content semantics. In this paper, we propose Decoupled Multimodal Fusion (DMF), a novel framework that introduces modality-enriched modeling to enable dynamic, context-aware fusion of ID-based and multimodal representations during user interest encoding. We design Decoupled Target Attention (DTA), which avoids redundant computation by reusing target-agnostic ID-derived components and integrating similarity signals via discretized embedding lookup. Furthermore, DMF unifies modality-centric and modality-enriched pathways through Complementary Modality Modeling (CMM) to balance semantic generalization and behavioral personalization. Offline experiments on public and industrial datasets demonstrate the effectiveness of DMF. More importantly, under industrial-scale serving conditions with 1200 candidate items, DTA delivers three times higher throughput compared to early fusion, enabling deployment with negligible latency overhead. DMF has been successfully deployed on Lazada's product recommendation system, yielding significant improvements of +5.30\% in CTCVR and +7.43\% in GMV.
\end{abstract}


\section{Introduction}
Click-Through Rate (CTR) prediction is a fundamental task in industrial recommender systems, aiming to estimate the probability that a user will click on a candidate item. The predicted CTR serves as a crucial signal for ranking decisions~\cite{deepfm2017,miss2022,attention2023,him2023}. User interest modeling, which focuses on capturing evolving user preferences from historical interaction sequences, has become a key component of modern CTR prediction models and has contributed to notable performance improvements~\cite{din2018,transact2023}. However, most industrial systems still rely solely on sparse ID-based features, which often lead to insufficient item representation learning, thereby degrading CTR prediction performance. In addition, ID-based embeddings effectively capture behavioral collaboration from user interactions, yet fail to represent the rich semantic information embedded in multimodal content.

Recent advances in multimodal foundation models have inspired efforts to inject semantic knowledge into recommendation systems~\cite{gonext2023,universal2022}. As online computing resources are limited and response time constraints are strict, industrial systems typically adopt a two-stage framework to incorporate multimodal information into ID-based CTR prediction models~\cite{make2024}. The framework first extracts multimodal representations from pre-trained models and then integrates them into the downstream ID-based model while keeping the representations frozen. In the integration stage, directly incorporating multimodal representations into ID-based models suffers from the semantic gap across different embedding spaces, leading to suboptimal performance~\cite{adrecom2024}.

\begin{figure*}[tbp]
  \centering
  \includegraphics[width=1.0\linewidth]{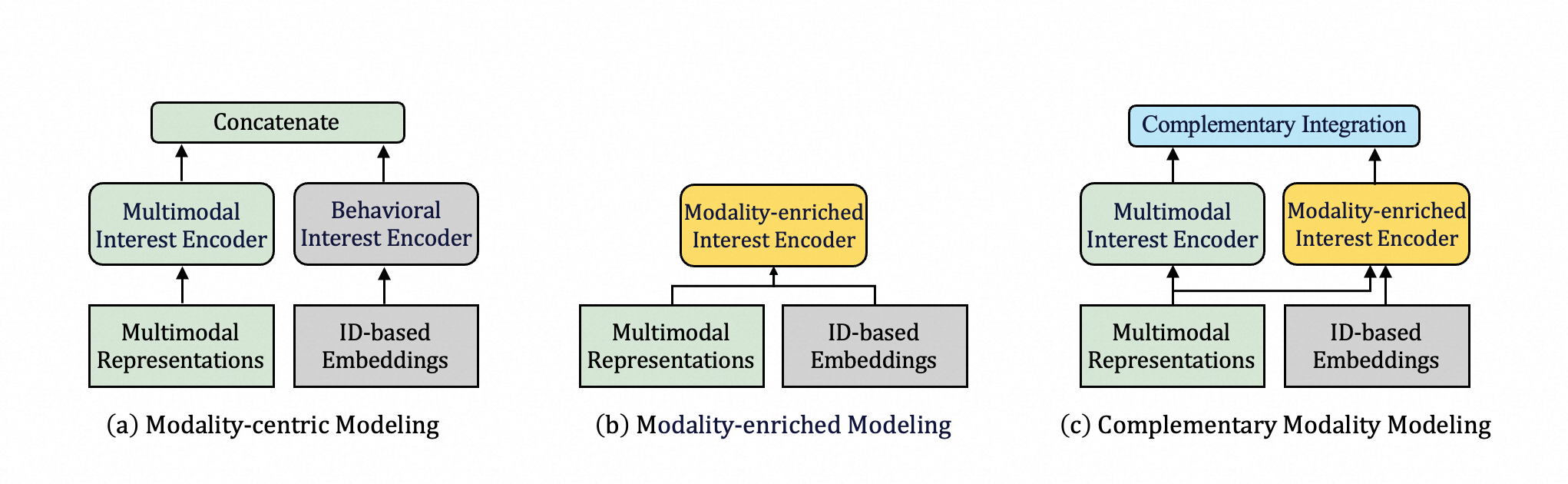}
  \caption{(a) Modality-centric Modeling: ID-based embeddings and multimodal representations are encoded independently, without fine-grained interaction during user interest modeling. (b) Modality-enriched Modeling: Multimodal representations are integrated as side information into the interest encoder, enabling fine-grained interaction between content semantics and behavioral signals during user interest modeling. (c) Complementary Modality Modeling (CMM): A structured fusion strategy that combines modality-centric and modality-enriched pathways. This hybrid approach produces a comprehensive user interest representation balancing semantic generalization and fine-grained behavioral personalization.}
  \label{fig:first-fig}
\end{figure*}

To mitigate the misalignment, recent approaches leverage multimodal similarity relationships between target items and historical interaction sequences as proxies for aligning embedding spaces, resulting in improved recommendation performance~\cite{make2024,distri2025}. The similarity histogram-based approach~\cite{make2024} converts high-dimensional multimodal representations into a compact similarity histogram by partitioning pairwise similarity scores into several discrete tiers and counting the number of matches in each bucket. The attention-based approach~\cite{distri2025} uses encoded similarity embeddings as item representations for behavioral-level multimodal interest fusion. However, both approaches adopt modality-centric modeling strategies (Fig.~\ref{fig:first-fig}(a)) that process ID-based embeddings and multimodal representations independently, falling short in capturing fine-grained interactions between content semantics and behavioral signals.

When seeking to enable fine-grained interactions between ID-based collaborative representations and multimodal representations, one might naturally consider treating multimodal similarity scores as side information and feeding them together with ID-based embeddings into a user interest modeling module, such as the Target Attention (TA) network~\cite{tin2024}. Although this modality-enriched modeling (Fig.~\ref{fig:first-fig}(b)) approach is straightforward, it faces two challenges when deployed in large-scale industrial recommendation systems: First, multimodal similarity scores are target-aware features, which means that different candidate items within the same request yield distinct similarity score sequences, even when derived from an identical historical interaction sequence. Using such features as side information requires recomputing the key ($K$) and value ($V$) representations for each candidate during online inference, making it impossible to reuse $K$ and $V$~\cite{twin2023}. The prohibitive computational overhead is unacceptable for industrial recommendation systems with more than $10^3$ candidate items~\cite{eta2021,sdim2022}. Second, even if computationally feasible, relying solely on fusing ID-based collaborative representations and multimodal representations in user interest modeling may lead to underutilization of multimodal information, as behavioral-level fusion is insufficient to capture a holistic representation of user multimodal interests.

To address these challenges, we propose \textbf{D}ecoupled \textbf{M}ultimodal \textbf{F}usion (\textbf{DMF}), a hybrid framework that introduces a modality-enriched modeling strategy to enable fine-grained interactions between ID-based collaborative representations and multimodal representations for user interest modeling, and combines modality-centric and modality-enriched paradigms to leverage their complementary strengths for boosting CTR prediction performance. For the first challenge, we propose Decoupled Target Attention (DTA) to integrate target-aware features into user interest modeling without sacrificing inference efficiency. For the second challenge, to fully exploit multimodal signals, we propose Complementary Modality Modeling (CMM)(Fig.~\ref{fig:first-fig}(c)), which fuses user interest representations learned under modality-centric and modality-enriched frameworks.

Our contributions can be summarized as follows:
\begin{itemize}
\item We identify the limitations of existing modality-centric approaches in capturing fine-grained semantics and propose modality-enriched modeling as a new paradigm that dynamically integrates multimodal signals into behavioral sequence encoding.
\item We propose Decoupled Target Attention (DTA), a new attention architecture that decouples ID and multimodal feature processing to achieve both expressive power and inference efficiency. DTA avoids redundant computation by reusing target-agnostic components, making it scalable to industrial-scale candidate sets.
\item We design Complementary Modality Modeling (CMM) to unify the strengths of modality-centric modeling and modality-enriched modeling strategies, achieving both semantic generalization and behavioral personalization. Extensive experiments demonstrate that DMF outperforms state-of-the-art methods and has been successfully deployed in the product recommendation system of the international e-commerce platform Lazada, yielding significant gains in CTCVR (+5.30\%) and GMV (+7.43\%).
\end{itemize}

\section{Related Work}\label{sec:rw}

In this section, we review related work in three major areas, i.e., click-through rate prediction, multimodal recommendation, and side information fusion.

\subsection{Click-through Rate Prediction}

CTR prediction plays an important role in modern recommendation systems, aiming to estimate the likelihood of a user clicking a candidate item~\cite{clickprompt2024,discrete2024}. Approaches like DeepFM~\cite{deepfm2017} and DCN~\cite{dcn2017} are designed for modeling interactions. Sequential approaches focus on modeling user interests from historical interaction sequences, which can be roughly classified into target-agnostic approaches and target-aware approaches. Approaches like SASRec~\cite{sasrec2018} and BERT4Rec~\cite{bert4rec2019} are formulated in a target-agnostic fashion, where user interaction sequences are independent of the candidate item and the derived user representations remain fixed across different candidate items. In contrast, most mainstream CTR prediction approaches~\cite{twin2023,din2018,sdim2022,transact2023,apic2024} used in industrial settings are formulated in a target-aware fashion that adaptively models user interests by considering the relevance between candidate items and historically interacted items. However, these approaches rely solely on sparse ID-based features, which fail to represent the rich semantic information embedded in multimodal content, thereby degrading CTR prediction performance.

\subsection{Multimodal Recommendation}

The rich semantic information embedded in multimodal data has spurred extensive efforts to incorporate it into recommendation systems. The potential of learning multimodal representations in an end-to-end manner alongside recommendation model training has been investigated in prior work~\cite{gonext2023}. Despite their effectiveness, these approaches incur high computational costs, rendering them impractical for large-scale industrial deployment. Recently, a two-stage framework~\cite{adrecom2024,pinnersage2020} has been introduced in industrial systems to leverage multimodal data for both effectiveness and cost efficiency. Building on this success, some studies have adopted the well-established two-stage paradigm and focus on improving the integration stage~\cite{make2024,distri2025}. 

Since directly incorporating multimodal representations into ID-based models is ineffective due to distinct spaces~\cite{cour2025,diffusion2025}, a variety of approaches have been proposed to address this misalignment. MARN~\cite{gan2020} draws inspiration from Generative Adversarial Networks (GANs), leveraging an adversarial mechanism to achieve alignment across modalities. MAKE~\cite{make2024} proposes a three-stage framework that decouples the optimization of multimodal related parameters from that of ID-based features via a recommendation pre-training module. By encoding similarity scores into learnable embeddings, DMAE~\cite{distri2025} aligns heterogeneous semantic spaces without requiring direct compatibility between raw embedding vectors. As similarity encoding avoids the instability and complexity of adversarial or multi-stage alignment by distilling relational knowledge from external embeddings into a learnable space, in this paper, we employ multimodal similarity scores between target items and historically interacted items as alignment proxies and side information to enhance the effectiveness of user interest modeling.

Within the similarity-based alignment framework, existing multimodal fusion approaches can be broadly categorized into similarity histogram-based modeling and attention-based modeling. Similarity histogram-based modeling~\cite{make2024} calculates a statistical similarity histogram for user interest modeling. Attention-based modeling~\cite{distri2025} utilizes the encoded similarity embeddings as input representations for behavioral-level multimodal interest fusion. However, both approaches process ID-based embeddings and multimodal representations independently, failing to capture fine-grained interactions between content semantics and behavioral signals. This limitation motivates the development of a more expressive multimodal fusion approach.

\subsection{Side Information Fusion}

In sequential recommendation, side information fusion approaches integrate various item attributes into the overall item representation, enabling the model to capture diverse collaborative signals across items~\cite{sideinfosigir2025}. From the perspective of when feature representations are merged, existing self-attention-based side information fusion approaches can be classified into early, late, and hybrid fusion~\cite{sideinfowww2024}. 

Early fusion~\cite{s3rec2020} integrates item ID and side information before feeding them into the model, facilitating interaction between item ID and side information representations. However, this simple aggregation may lead to information invasion~\cite{nova2011} due to inherent differences in representation spaces. In contrast, late fusion postpones merging item ID and side information until the final prediction stage. For example, FDSA~\cite{fdsa2019} uses independent self-attention blocks to model item ID and side information in isolation, thereby preserving their distinct contextual patterns. Although such isolation supports accurate modeling within each modality, it inherently limits the ability to learn fine-grained interactions between item ID and side information. Hybrid fusion lies between early and late fusion, allowing item ID and side information to interact in the middle layer. NOVA~\cite{nova2011} proposes only to incorporate attributes in the calculation of attention scores to mitigate the information invasion problem. DIF-SR~\cite{difsr2022} decouples the attention calculation of various side information and item ID representation, allowing higher-rank attention matrices and flexible gradients. ASIF~\cite{sideinfowww2024} utilizes side information without noisy interference via fused attention with untied position information. 

However, these side information fusion approaches are designed for target-agnostic sequential modeling, which is based on inherent item attributes that are independent of the candidate item. Directly applying these approaches to target-aware side information fusion may result in suboptimal performance or prohibitive computational overhead. This calls for a framework tailored to target-aware side information fusion that enables competitive recommendation performance while preserving online inference efficiency.

\section{Preliminaries}\label{sec:pr}

\subsection{Task Formulation}

The goal of CTR prediction is to estimate the probability that a user clicks a candidate item, referred to as the \textbf{target item}. Formally, let $\mathcal{D} = \{(x_i, y_i)\}$ denote the training set, where $x_i$ represents the input feature vector of the $i$-th instance and $y_i \in \{0, 1\}$ is the binary label indicating whether a click occurs ($y_i = 1$) or not ($y_i = 0$). The predicted CTR is computed as:
\begin{equation}
    \hat{y}_i = \mathcal{F}(x_i),
    \label{eq:ctr_prediction}
\end{equation}
where $\mathcal{F}(\cdot)$ denotes the CTR prediction model, which includes an embedding layer mapping categorical features into dense vectors.

The model is trained by minimizing the binary cross entropy loss:
\begin{equation}
    \mathcal{L} = -\frac{1}{|\mathcal{D}|} \sum_{i=1}^{|\mathcal{D}|} \left( y_i \log \hat{y}_i + (1 - y_i) \log(1 - \hat{y}_i) \right).
    \label{eq:bce_loss}
\end{equation}

\begin{figure*}[tbp]
  \centering
  \includegraphics[width=1.0\linewidth]{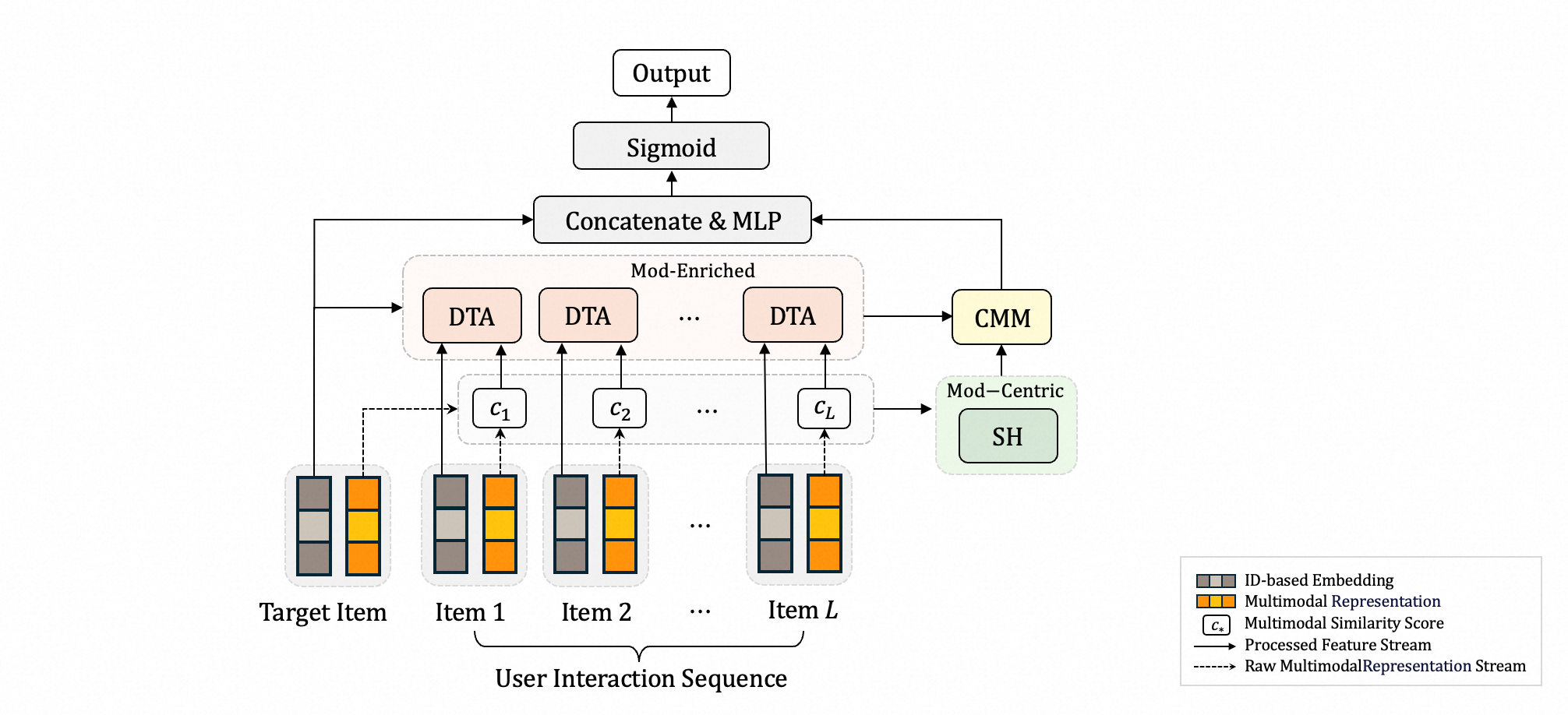}
  \caption{The framework of DMF. Multimodal representations are used to compute similarity scores between the target item and each historically interacted item. These scores are discretized and mapped into learnable embeddings as input to the DTA module, while SH (Similarity Histogram) operates directly on the raw similarity scores. Mod-Centric and Mod-Enriched denote the Modality-Centric and Modality-Enriched modeling strategies, respectively. The resulting user interest representations are fused by the CMM module and combined with the target item for CTR prediction.}
  \label{fig:dmf}
\end{figure*}

\subsection{Target-aware Attention}\label{sec:ta}

User interest modeling often employs target-aware attention mechanisms to adaptively capture relevant historical interactions with respect to the target item~\cite{din2018,twin2023,sdim2022,tin2024}. In this framework, for a given user interaction sequence of length $L$, let $\mathbf{s}_j \in \mathbb{R}^d$ denote the embedding of the $j$-th historically interacted item for $j = 1, \dots, L$. Let $\mathbf{q} \in \mathbb{R}^{d_q}$ denote the embedding of the target item. The sequence embeddings are used to construct the key matrix $\mathbf{K} \in \mathbb{R}^{L \times d_k}$ and value matrix $\mathbf{V} \in \mathbb{R}^{L \times d_v}$, while the target embedding serves as the query $\mathbf{Q} \in \mathbb{R}^{1 \times d_q}$. In mini-batch training, these are extended to $\mathbf{K} \in \mathbb{R}^{B \times L \times d_k}$, $\mathbf{V} \in \mathbb{R}^{B \times L \times d_v}$, and $\mathbf{Q} \in \mathbb{R}^{B \times d_q}$, where $B$ is the batch size, $L$ is the maximum sequence length, and $d_k$, $d_v$, $d_q$ are the dimensions of key, value, and query vectors, respectively.

For the $i$-th sample in the batch, let $\mathbf{q}_i \in \mathbb{R}^{d}$ denote the query vector corresponding to the target item. Let $\mathbf{s}_j \in \mathbb{R}^d$ denote the embedding of the $j$-th historically interacted item for $j = 1, \dots, L$. The historical embedding matrix is defined as $\mathbf{S} = [\mathbf{s}_1, \mathbf{s}_2, \ldots, \mathbf{s}_L]^\top \in \mathbb{R}^{L \times d}$. The attention mechanism computes a user interest representation $\mathbf{u}_i \in \mathbb{R}^{d_v}$ as a weighted sum over the value vectors:
\begin{equation}
    \mathbf{u}_i = \sum_{j=1}^{L} \alpha_{ij} \mathbf{v}_j,
    \label{eq:weighted_sum}
\end{equation}
where $\mathbf{v}_j = \mathbf{V}_{i,j,:} \in \mathbb{R}^{d_v}$ denotes the $j$-th value vector in the $i$-th sample, and $\alpha_{ij}$ is the attention weight assigned to the $j$-th item. The weights are computed via softmax normalization over unnormalized alignment scores $e_{ij} \in \mathbb{R}$:
\begin{equation}
    \alpha_{ij} = \frac{\exp(e_{ij} / \tau)}{\sum_{k=1}^{L} \exp(e_{ik} / \tau)},
    \label{eq:softmax_attention}
\end{equation}
where $\tau$ is a temperature parameter~\cite{attention2017}.

DIN~\cite{din2018} and MHTA~\cite{twin2023} are the two most representative
target-aware attention approaches. The main difference between them is how the score $e_{ij}$ is obtained. In DIN~\cite{din2018}, the key and value vectors are set to the input embeddings of the historically interacted items:
\begin{equation}
    \mathbf{k}_j = \mathbf{s}_j, \quad \mathbf{v}_j = \mathbf{s}_j,
\end{equation}
and $e_{ij}$ is computed as:
\begin{equation}
    e_{ij} = \mathrm{MLP}\left([\mathbf{q}_i; \mathbf{s}_j; \mathbf{q}_i \odot \mathbf{s}_j; \mathbf{q}_i - \mathbf{s}_j]\right),
\end{equation}
where $[\cdot;\cdot]$ denotes vector concatenation, $\odot$ denotes element-wise multiplication, and $\mathrm{MLP}$ is a multilayer perceptron with PReLU~\cite{prelu2015} activations. This design explicitly models both similarity and dissimilarity between the target item and historical interactions.

MHTA~\cite{twin2023} computes keys and values via linear projections. Let $\mathbf{W}_q, \mathbf{W}_k, \mathbf{W}_v \in \mathbb{R}^{d_h \times d}$ be learnable weight matrices, where $d_h$ is the attention hidden dimension, which corresponds to the key and value dimensions, i.e., $d_k = d_v = d_h$. The projected key and value vectors are:
\begin{align}
    \mathbf{k}_j &= \mathbf{W}_k \mathbf{s}_j \in \mathbb{R}^{d_h}, \\
    \mathbf{v}_j &= \mathbf{W}_v \mathbf{s}_j \in \mathbb{R}^{d_h},
\end{align}
and $e_{ij}$ is computed using scaled dot-product attention~\cite{attention2017}:
\begin{equation}
    e_{ij} = \frac{(\mathbf{W}_q \mathbf{q}_i)^\top \mathbf{k}_j}{\sqrt{d_h}},
    \label{eq:mhta_score}
\end{equation}
where $\mathbf{W}_q \mathbf{q}_i \in \mathbb{R}^{d_h}$ serves as the query vector.

Although both approaches follow the same attention paradigm, they differ in the construction of $\mathbf{K}$, $\mathbf{V}$, and the computation of $e_{ij}$. 
In DIN, $e_{ij}$ depends nonlinearly on both $\mathbf{q}_i$ and $\mathbf{s}_j$, making it expressive but preventing reuse of intermediate representations across different target items. As a result, the entire attention module must be recomputed for each target item during inference.

In MHTA, the key and value matrices $\mathbf{K} = \mathbf{S} \mathbf{W}_k^\top$ and $\mathbf{V} = \mathbf{S} \mathbf{W}_v^\top$ are functions of the historical sequence only and do not depend on the target item. As a result, once computed for a given user sequence, they can be reused across multiple target items without recomputation. This structural property reduces redundant computation and improves inference efficiency in industrial recommendation systems.

The computational bottleneck in DIN arises from the MLP applied to each query-key pair, while in MHTA it stems from the linear projections of the key and value vectors~\cite{sdim2022,twin2023}. Our proposed Decoupled Target Attention (DTA), introduced in Section~\ref{sec:dta}, builds upon this foundation to support expressive multimodal fusion while preserving inference efficiency.

\section{Methodology}\label{sec:method}

The framework of DMF is illustrated in Fig.~\ref{fig:dmf}. 
Due to the semantic gap between ID-based embeddings and multimodal representations, we use target-aware similarity scores as alignment signals. These features are fed into both the Similarity Histogram and DTA modules to derive complementary interest representations. In particular, the constructed features are transformed into learnable embeddings to enable computational decoupling within the DTA module. The CMM module fuses these representations to capture both semantic generalization and fine-grained behavioral patterns. The final representation is concatenated with the ID embedding of the target item and fed into an MLP for prediction.

\subsection{Target-aware Multimodal Similarity Feature}\label{sec:feature}

Images and text descriptions capture rich semantic signals that reflect item relationships, which we leverage as multimodal inputs. We use RoBERTa~\cite{roberta2019} and Vision Transformer (ViT)~\cite{vit2020} to encode item titles and main images into textual and visual representations, respectively. These representations are fused and used to fine-tune the encoders via contrastive learning on item pairs, encouraging semantically similar items to have aligned embeddings. The resulting frozen embeddings are used to compute multimodal similarity scores in downstream modeling.

As previously discussed, direct feature fusion with raw embeddings of multimodal information is ineffective for CTR prediction due to distinct spaces. State-of-the-art multimodal recommenders often freeze pretrained modality encoders and leverage item similarity computed by multimodal representations as auxiliary input to guide user interest modeling. Let $\mathbf{m}_c \in \mathbb{R}^{d_m}$ and $\mathbf{m}_j \in \mathbb{R}^{d_m}$ denote the frozen multimodal representations of the target item and the $j$-th historically interacted item, respectively, where $d_m$ is the dimension of the multimodal representation. The multimodal similarity score between them is computed using cosine similarity:
\begin{equation}
    c_j = \frac{\mathbf{m}_c^\top \mathbf{m}_j}{\|\mathbf{m}_c\|_2 \|\mathbf{m}_j\|_2}, \quad \forall j \in \{1, \dots, L\}.
    \label{eq:cosine_sim}
\end{equation}
We define the similarity vector as $\mathbf{c}_{\text{sim}} = [c_1, c_2, \ldots, c_L]^\top \in \mathbb{R}^L$, where each $c_j \in [-1, 1]$ denotes the cosine similarity between the target item and the $j$-th historically interacted item. This vector serves as target-aware side information in subsequent modeling.

\subsection{Target-Aware and Target-Agnostic Nodes}

\begin{figure}[tbp]
  \centering
  \includegraphics[width=1.0\linewidth]{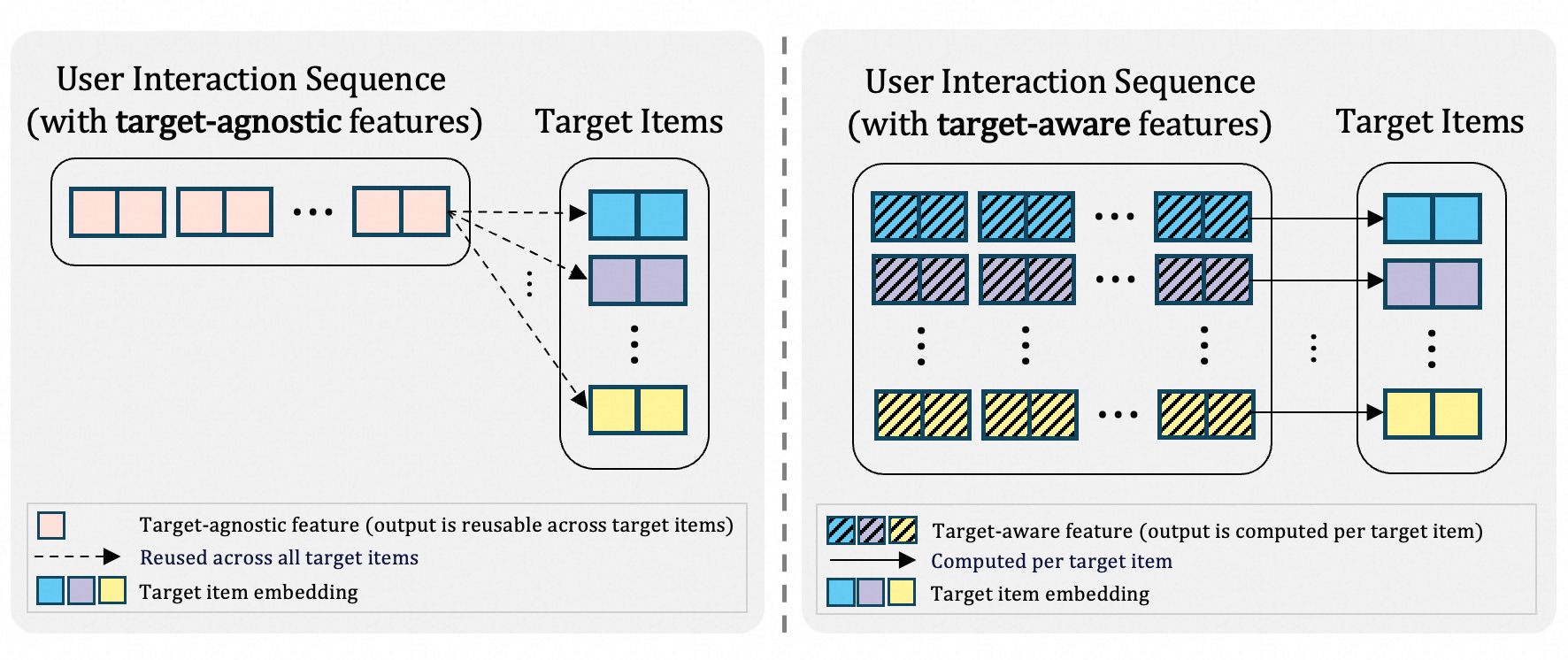}
  \caption{Illustration of target-agnostic and target-aware node computation in a CTR model. Left panel shows target-agnostic features derived from user interaction sequence, which are computed once and reused across all target items. Right panel shows target-aware features, where each target item triggers a distinct computation path for its associated historical interactions. The output of target-agnostic nodes is reusable, while target-aware nodes require recomputation per target item.}
  \label{fig:node}
\end{figure}

In the computation graph of a CTR model, we classify intermediate nodes into two categories based on their input features and tensor shapes, i.e., target-aware nodes and target-agnostic nodes. As illustrated in Fig.~\ref{fig:node}, this distinction has direct implications for online inference efficiency.

A target-aware node processes target-aware features for a batch of target items, and thus has a leading dimension of $B$, where $B$ denotes the number of items being scored. In contrast, a target-agnostic node processes user-level features (e.g., demographics, historically interacted items) and outputs a $1 \times d$ tensor computed once per user. For example, in a 2-dimensional tensor representation, a target-agnostic node outputs a tensor of size $1 \times d$, while an item-side node outputs $B \times d$. Due to the shape difference, target-agnostic nodes are often more computationally efficient, as the computational overhead does not scale with the number of target items\footnote{When a downstream operator (e.g., concatenation or element-wise addition) takes inputs from both a target-agnostic node and a target-aware node, broadcast-compatible operators can be used to align the shapes of these two nodes, avoiding redundant computation caused by explicit replication.}. The reusability property of target-agnostic nodes has been widely exploited in industrial systems to optimize online inference complexity~\cite{hstu2024}.

\subsection{Decoupled Target Attention}\label{sec:dta}

\begin{figure*}[tbp]
  \centering
  \includegraphics[width=1.0\linewidth]{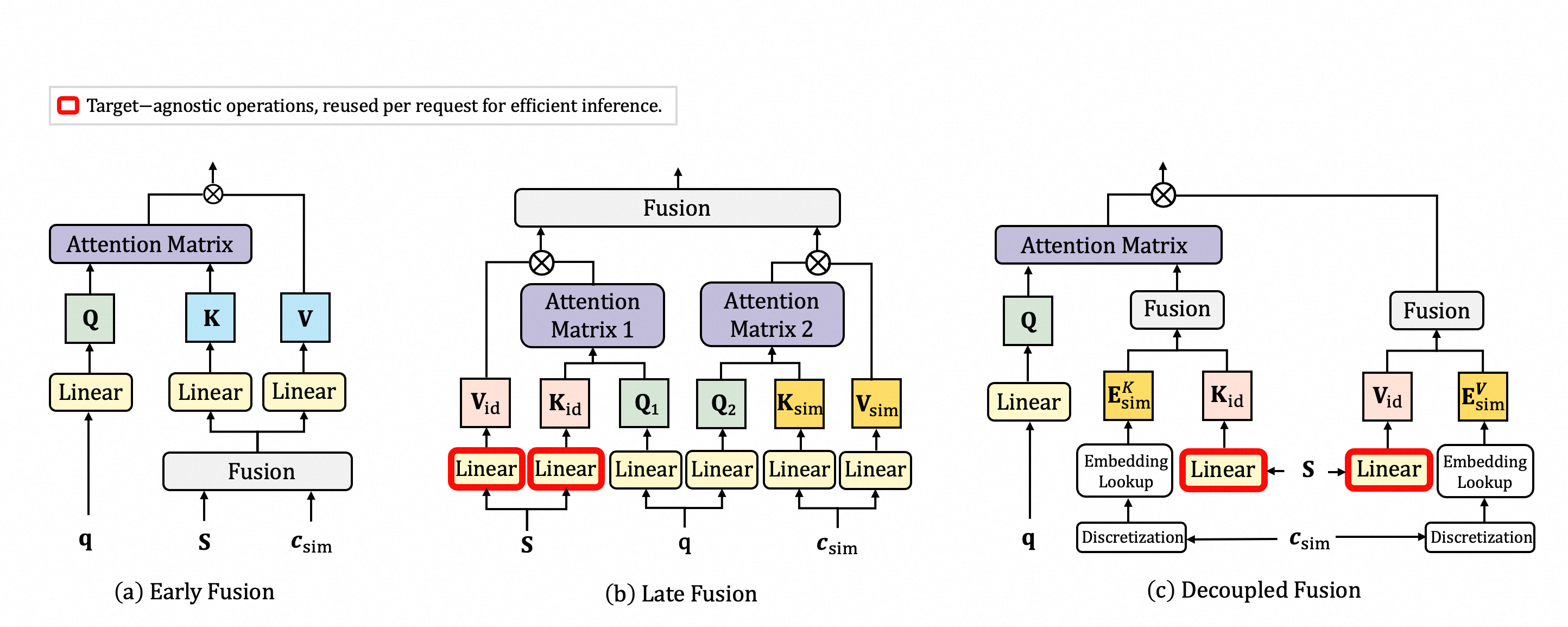}
  \caption{Comparison of side information fusion methods based on target-aware attention: (a) early fusion, (b) late fusion, and (c) decoupled fusion, which balances efficiency and effectiveness.}
  \label{fig:fusion}
\end{figure*}

Compared with DIN, the online inference complexity of MHTA can be greatly optimized by applying the reusability property of target-agnostic nodes to the linear projections of $\mathbf{K}$ and $\mathbf{V}$. DIN cannot benefit from this optimization due to the early fusion between the target items and sequential interactions before being fed into the MLP. Therefore, we employ MHTA as the target-aware attention approach and integrate the target-aware multimodal similarity feature constructed in Section~\ref{sec:feature} into MHTA.

A straightforward way for integrating the target-aware multimodal similarity vector $\mathbf{c}_{\text{sim}} = [c_1, c_2, \ldots, c_L]^\top$ is to treat it as side information and feed it together with ID-based embeddings into MHTA. As shown in Fig.~\ref{fig:fusion}(a), this approach is referred to as early fusion. By incorporating ID-based embeddings and similarity scores at the input level, the key matrix $\mathbf{K}$ and value matrix $\mathbf{V}$ become target-aware nodes. Prior work shows that target-aware computation tends to yield better model performance when integrated earlier into the sequential model structure~\cite{transact2023}. However, the prohibitive computational overhead is unacceptable for industrial recommendation systems with more than $10^3$ target items. Late fusion defers the integration of ID-based embeddings and similarity scores until the final prediction layer, as illustrated in Fig.~\ref{fig:fusion}(b). In late fusion, ID-based embeddings and multimodal similarity scores are processed separately, where the ID-based embeddings are projected via linear layers to obtain $\mathbf{K}_{\mathrm{id}}$ and $\mathbf{V}_{\mathrm{id}}$, while the similarity scores $\{c_j\}$ are similarly projected to produce auxiliary key and value matrices $\mathbf{K}_{\mathrm{sim}}$ and $\mathbf{V}_{\mathrm{sim}}$. The $\mathbf{K}_{\mathrm{id}}$ and $\mathbf{V}_{\mathrm{id}}$ computed by ID-based embeddings become target-agnostic nodes, and the calculations can be reused. As the late fusion method utilizes a modality-centric modeling strategy, it fails to capture fine-grained interactions between content semantics and behavioral signals. All these weaknesses motivate us to find a fusion model for target-aware features that balances efficiency and effectiveness.

In this paper, we propose Decoupled Target Attention (DTA), which decouples the computation of the $\mathbf{K}$ based on ID-derived embeddings from that based on multimodal similarity features, as shown in Fig.~\ref{fig:fusion}(c). DTA extends MHTA by augmenting its key and value projections with modality-aware signals, while preserving the target-agnostic structure of MHTA for $\mathbf{K}_{\mathrm{id}}$ and $\mathbf{V}_{\mathrm{id}}$.
The same decoupling applies to the computation of the value matrix $\mathbf{V}$. For the ID-derived components, the key matrix $\mathbf{K}_{\mathrm{id}}$ and value matrix $\mathbf{V}_{\mathrm{id}}$ are target-agnostic and thus reusable across different target items. For the multimodal similarity features, each scalar similarity score $c_j$  is discretized into a bucket index $b_j = \mathrm{bucket}(c_j)$, and then mapped into separate learnable embedding vectors for keys and values, where $\mathbf{e}_{b_j}^K, \mathbf{e}_{b_j}^V \in \mathbb{R}^{d_h}$. The raw cosine similarities are processed by discretizing them into buckets and performing embedding lookup, enabling differentiable and efficient integration of target-aware side information. This avoids linear projection layers when computing auxiliary key $\mathbf{K}_{\mathrm{sim}}$ and value $\mathbf{V}_{\mathrm{sim}}$, which would otherwise necessitate $B$ times recomputation per user sequence during online inference.

To formalize this design, let $\mathbf{S} = [\mathbf{s}_1, \mathbf{s}_2, \ldots, \mathbf{s}_L]^\top \in \mathbb{R}^{L \times d}$ denote the ID-based embedding matrix of the historical interaction sequence, and let $\mathbf{c}_{\text{sim}} = [c_1, c_2, \ldots, c_L]^\top \in \mathbb{R}^L$ denote the multimodal similarity scores between the target item and each historical item, as defined in Eq.~\eqref{eq:cosine_sim}. Our goal is to compute the query $\mathbf{Q}^{\mathrm{DTA}} \in \mathbb{R}^{B \times d_h}$, key $\mathbf{K}^{\mathrm{DTA}} \in \mathbb{R}^{L \times d_h}$, and value $\mathbf{V}^{\mathrm{DTA}} \in \mathbb{R}^{L \times d_h}$ for DTA.

Specifically, each similarity score $c_j \in [-1, 1]$ is discretized into one of $M$ predefined buckets using equal frequency discretization~\cite{bucket2021}, producing a bucket index $b_j \in \{1,\dots,M\}$. Let $b_j = \text{bucket}(c_j)$ denote the bucket index of $c_j$. Each bucket $m \in \{1,\dots,M\}$ corresponds to two independent learnable embedding vectors $\mathbf{e}_m^K, \mathbf{e}_m^V \in \mathbb{R}^{d_h}$, forming two lookup tables $\{\mathbf{e}_m^K\}_{m=1}^M$ and $\{\mathbf{e}_m^V\}_{m=1}^M$. We define the similarity embedding matrices as:
\begin{align}
    \mathbf{E}_{\mathrm{sim}}^K &= [\mathbf{e}_{b_1}^K, \mathbf{e}_{b_2}^K, \ldots, \mathbf{e}_{b_L}^K]^\top \in \mathbb{R}^{L \times d_h}, \\
    \mathbf{E}_{\mathrm{sim}}^V &= [\mathbf{e}_{b_1}^V, \mathbf{e}_{b_2}^V, \ldots, \mathbf{e}_{b_L}^V]^\top \in \mathbb{R}^{L \times d_h}.
\end{align}

Instead of projecting similarity scores through linear layers (which would require $B$-times computation), we discretize scores into buckets and perform embedding lookup, a constant-time operation independent of $B$. Subsequently, $\mathbf{E}_{\mathrm{sim}}^K$ and $\mathbf{E}_{\mathrm{sim}}^V$ are added to $\mathbf{K}_{\mathrm{id}}$ and $\mathbf{V}_{\mathrm{id}}$, respectively, to form the final key and value matrices for scaled dot-product attention.

To ensure that the decoupled fusion strategy does not sacrifice expressive power compared to the early fusion approach, we formally establish a condition under which both methods are computationally equivalent. Specifically, we show that when the discretization scheme employs a coarse bucketing strategy with sufficient capacity, the output of decoupled fusion can approximate that of early fusion to arbitrary precision.

\medskip
\noindent\textbf{Theorem 1} (Approximate Expressiveness of Decoupled Fusion). {\itshape To illustrate the expressiveness of our discretization scheme, we consider a simplified early fusion model where the hidden representation of the $i$-th historical item is computed as:
\begin{equation}
    \mathbf{h}_i' = \mathbf{s}_i + c_i \cdot \mathbf{w},
\end{equation}
where $\mathbf{s}_i \in \mathbb{R}^d$ is the ID-based embedding of the $i$-th historically interacted item, $c_i \in [-1,1]$ denotes its multimodal similarity score with respect to the target item, and $\mathbf{w} \in \mathbb{R}^d$ is a shared learnable parameter vector that modulates the influence of the similarity signal.

In contrast, the proposed decoupled fusion computes:
\begin{equation}
    \mathbf{h}_i^{\text{final}} = \mathbf{s}_i + \mathbf{e}_{b_i},
\end{equation}
where $b_i = \text{Bucket}(c_i)$ maps $c_i$ to one of $M$ discrete buckets, and $\mathbf{e}_{b_i} \in \mathbb{R}^{d_h}$ is a learnable embedding associated with bucket $b_i$.

Then, for any arbitrarily small approximation error $\epsilon > 0$, there exists a sufficiently large number of buckets $M$ and an appropriate choice of $\{\mathbf{e}_m\}_{m=1}^M$ such that:
\begin{equation}
    \max_{i} \|\mathbf{h}_i' - \mathbf{h}_i^{\text{final}}\|_2 < \epsilon.
\end{equation}
}

Theorem 1 justifies our design choice. Through a coarse-bucketing scheme endowed with sufficient capacity, DTA attains nearly lossless expressiveness while delivering substantial computational efficiency. Despite its simplicity, the discretization with learnable embeddings effectively captures complex patterns in the similarity signal, preserving the modeling power of decoupled fusion.

Let $\mathbf{S} = [\mathbf{s}_1, \mathbf{s}_2, \ldots, \mathbf{s}_L]^\top \in \mathbb{R}^{L \times d}$ denote the ID-based embedding matrix of the historical interaction sequence. We compute the query, key, and value for Decoupled Target Attention (DTA) as follows. The query $\mathbf{Q}^{\mathrm{DTA}} \in \mathbb{R}^{B \times d_h}$ is derived from the target item embeddings $\mathbf{q}_i$ for each target item:
\begin{equation}
    \mathbf{q}_i^{\mathrm{DTA}} = \mathbf{W}_q \mathbf{q}_i,\quad
    \mathbf{Q}^{\mathrm{DTA}} = [\mathbf{q}_1^{\mathrm{DTA}}, \ldots, \mathbf{q}_B^{\mathrm{DTA}}] \in \mathbb{R}^{B \times d_h},
    \label{eq:q_d}
\end{equation}
where $\mathbf{W}_q \in \mathbb{R}^{d_h \times d}$ is a learnable projection matrix.

The base key and value matrices are computed from ID-based embeddings:
\begin{align}
    \mathbf{K}_{\mathrm{id}} &= \mathbf{S} \mathbf{W}_k^\top \in \mathbb{R}^{L \times d_h}, \\
    \mathbf{V}_{\mathrm{id}} &= \mathbf{S} \mathbf{W}_v^\top \in \mathbb{R}^{L \times d_h},
\end{align}
where $\mathbf{W}_k, \mathbf{W}_v \in \mathbb{R}^{d_h \times d}$ are learnable projection matrices. These matrices are independent of the target item and thus reusable across multiple target items.

The final key and value matrices are obtained by element-wise addition:
\begin{align}
    \mathbf{K}^{\mathrm{DTA}} &= \mathbf{K}_{\mathrm{id}} + \mathbf{E}_{\mathrm{sim}}^K, \\
    \mathbf{V}^{\mathrm{DTA}} &= \mathbf{V}_{\mathrm{id}} + \mathbf{E}_{\mathrm{sim}}^V,
\end{align}
ensuring fine-grained interaction while preserving reusability of $\mathbf{K}_{\mathrm{id}}$ and $\mathbf{V}_{\mathrm{id}}$.

The output of the DTA module is computed as:
\begin{equation}
    \mathbf{U}^{\mathrm{DTA}} = \mathrm{DTA}(\mathbf{Q}^{\mathrm{DTA}}, \mathbf{K}^{\mathrm{DTA}}, \mathbf{V}^{\mathrm{DTA}}) \in \mathbb{R}^{B \times d_h},
\end{equation}
where $\mathbf{U}^{\mathrm{DTA}} \in \mathbb{R}^{B \times d_h}$ is the resulting user interest representation.

\subsection{Complementary Modality Modeling}

The modality-centric and modality-enriched modeling strategies capture user interests from fundamentally different perspectives, making them complementary. The modality-centric modeling strategy answers whether an item is semantically consistent with user interests, supporting broad generalization, while the modality-enriched modeling strategy determines which among the semantically related items has been most behaviorally significant, enabling precise personalization. Their integration allows for benefiting from both stable semantic matching and fine-grained interaction modeling.

For modality-centric modeling, we employ a similarity histogram-based approach~\cite{make2024} to obtain user interest representations. Given the multimodal similarity scores, the score range from $-1.0$ to $1.0$ is uniformly partitioned into $N$ predefined intervals. For each interval, the number of similarity scores falling within that range is counted. As a result, an $N$-dimensional histogram vector is constructed, where each dimension corresponds to the frequency of scores in the respective bucket. Then, the $N$-dimensional histogram vector is fed into a subsequent multilayer perceptron for further processing. For modality-enriched modeling, we exploit DTA to obtain user interest representations, detailed in Section~\ref{sec:dta}.

The final representation $\mathbf{r}_u$ is obtained by aggregating the user interest representations from modality-centric and modality-enriched modeling, respectively, as shown in Eq.~\eqref{eq:cmm}. While modality-centric modeling strategies emphasize semantic coherence through content-driven generalization, independent of item identities, modality-enriched modeling strategies capture fine-grained behavioral preferences by integrating identity signals with multimodal context, thereby prioritizing context-aware personalization.

Let $\mathbf{r}_{\mathrm{me}} = \mathbf{U}^{\mathrm{DTA}}_i \in \mathbb{R}^{d_h}$ denote the modality-enriched user interest representation from DTA, where $\mathbf{U}^{\mathrm{DTA}}_i$ is the $i$-th row of $\mathbf{U}^{\mathrm{DTA}}$, and let $\mathbf{r}_{\mathrm{mc}} \in \mathbb{R}^{d_h}$ denote the modality-centric representation derived from the similarity histogram through an MLP. The final user representation $\mathbf{r}_u \in \mathbb{R}^{d_h}$ is obtained via a convex combination of the two complementary representations:
\begin{equation}
    \mathbf{r}_u = \alpha \mathbf{r}_{\mathrm{me}} + (1 - \alpha) \mathbf{r}_{\mathrm{mc}},
    \label{eq:cmm}
\end{equation}
where $\alpha \in [0, 1]$ controls the trade-off between fine-grained personalization and semantic generalization.

\subsection{Discussion}

\textbf{Relationship with KV Caching.}
DTA is orthogonal to existing KV caching strategies~\cite{hstu2024} and addresses a distinct challenge in efficient attention computation. Conventional KV caching assumes that key and value representations are target-agnostic, which enables cache and reuse across target items. However, direct integration of target-aware features renders the resulting key and value matrices target-aware, thereby invalidating caching. DTA overcomes this limitation via structural decoupling. It isolates the target-agnostic ID-based pathway, which yields $\mathbf{K}_{\mathrm{id}}$ and $\mathbf{V}_{\mathrm{id}}$, from the target-aware multimodal enhancement. By encoding multimodal similarity scores through discretized embedding lookup rather than instance-specific linear projections, DTA preserves the reusability of base attention components while retaining expressive behavior-semantics interaction within the attention layer. Consequently, DTA does not merely coexist with KV caching. It enables effective caching in modality-enriched sequential recommendation, a regime where standard fusion approaches fail to support computational reuse.

\textbf{Computational Complexity Comparison of Fusion Strategies.}
To further quantify the efficiency advantage of DTA, we analyze the floating-point operations (FLOPs) required for computing the key ($\mathbf{K}$) and value ($\mathbf{V}$) matrices, which are the most computationally intensive part of attention~\cite{sdim2022,twin2023}. We assume the number of target items is $B$, the user interaction sequence length is $L$, and the embedding dimension before projection is $d_m$, with attention hidden size $d_h$. For simplicity and fair comparison, we set $d_m = d_h = d$. As shown in Table~\ref{tab:flops}, we compare three fusion strategies: early fusion, late fusion, and our proposed decoupled fusion, under two settings: (1) leveraging the reusability property of target-agnostic nodes, and (2) with such optimization enabled. The FLOPs are derived from the linear projections $\mathbf{W}_k, \mathbf{W}_v \in \mathbb{R}^{d \times d}$, where each projection of an $L \times d$ matrix costs $2Ld^2$ FLOPs (accounting for multiply-add operations).

\begin{table}[tbp]
\centering
\caption{FLOPs of TA leveraging the reusability property of target-agnostic nodes or not for $\mathbf{K}$ and $\mathbf{V}$ computation in different fusion strategies. We assume $B \gg L$ and $d_m = d_h = d$.}
\label{tab:flops}
\begin{tabular}{lcc}
\toprule
\textbf{Fusion Strategy} & \textbf{Reuse} & \textbf{No Reuse} \\
\midrule
Early Fusion      & $2BLd^2 + 2BLd$ & $2BLd^2 + 2BLd$ \\
Late Fusion       & $2Ld^2 + 2BLd$ & $2BLd^2 + 2BLd$ \\
Decoupled Fusion  & $2Ld^2 + 2BLd$ & $2BLd^2 + 2BLd$ \\
\bottomrule
\end{tabular}
\end{table}

In early fusion, both ID-based embeddings and similarity scores are fused before projection, rendering the resulting $\mathbf{K}$ and $\mathbf{V}$ fully target-aware. Consequently, no component can be reused across target items, and the FLOPs remain identical under both Reuse and No Reuse settings. In contrast, late fusion separates modality processing: the ID-based projections $\mathbf{K}_{\mathrm{id}}$ and $\mathbf{V}_{\mathrm{id}}$ are target-agnostic and thus reusable, reducing the dominant $2BLd^2$ term to $2Ld^2$ when reuse is enabled.

Our decoupled fusion achieves the same computational efficiency as late fusion under the Reuse setting—since $\mathbf{K}_{\mathrm{id}}$ and $\mathbf{V}_{\mathrm{id}}$ are computed once—while avoiding linear projections on the target-aware similarity scores altogether. Instead, similarity scores are discretized and mapped via embedding lookup, which incurs negligible computational cost (linear in $L$ and independent of $B$) and is omitted in the FLOPs count as it is dominated by projection terms. Therefore, the optimized FLOPs for decoupled fusion match those of late fusion, but with richer cross-modality interaction preserved within the attention mechanism.

To illustrate the practical impact, consider a realistic industrial scenario with $B = 1000$ candidate items, $L = 400$ historical interactions, and $d = 128$. The FLOPs for $\mathbf{K}$/$\mathbf{V}$ computation are:
\begin{itemize}
    \item \textbf{Early Fusion}: $2 \times 1000 \times 400 \times 128^2 + 2 \times 1000 \times 400 \times 128 \approx 13.10$ GFLOPs (identical under both reuse settings).
    \item \textbf{Late Fusion \& Decoupled Fusion (with reuse)}: $2 \times 400 \times 128^2 + 2 \times 1000 \times 400 \times 128 \approx 0.13$ GFLOPs.
\end{itemize}
This represents a \textbf{100$\times$ reduction} in FLOPs when reuse is enabled for late and decoupled fusion strategies, highlighting that DTA enables industrial-scale deployment without sacrificing modeling expressiveness.

\section{Experiments}\label{sec:exp}

In this section, we conduct extensive experiments to answer the following research questions:
\begin{itemize}
\item \textbf{RQ1}: How does the proposed DMF framework perform compared with the state-of-the-art CTR prediction baselines?
\item \textbf{RQ2}: How do different side information fusion strategies affect model performance and inference efficiency?
\item \textbf{RQ3}: How sensitive is DMF to the representation aggregation hyperparameter $\alpha$?
\item \textbf{RQ4}: How does the interaction sequence length influence the optimal trade-off between modality-centric and modality-enriched modeling?
\item \textbf{RQ5}: Can DMF capture meaningful user interest patterns in real-world scenarios?
\item \textbf{RQ6}: How does the proposed DMF framework perform on the real-world e-commercial platform?
\end{itemize}

\begin{table*}[tbp]
  \caption{Model Comparison on Amazon Dataset and Industrial Dataset. We report the mean and standard deviation (std) over 5 runs. $\Delta_{\text{AUC}}$ is calculated to indicate the average performance boost compared with the baseline (SASRec) over datasets. The best and second-best results are marked in bold and underline, respectively. For the industrial dataset, an improvement of 0.1\% in AUC and GAUC in the offline evaluation is considered significant for CTR prediction, as observed in~\cite{dcnv2021,twin2023}.}
  \label{tab:all_performance}
  \centering
  \renewcommand{\arraystretch}{1.1}
  \begin{tabular}{ccccccc}
    \toprule
    \multirow{3}{*}[+0.7ex]{\textbf{Model}} 
      & \multicolumn{2}{c}{\textbf{Amazon}} 
      & \multicolumn{4}{c}{\textbf{Industry}} \\
    \cmidrule(lr){2-3} \cmidrule(lr){4-7}
      & \textbf{\textit{AUC (mean $\pm$\text{ std})}} & \textbf{\textit{$\Delta_{\text{AUC}}\uparrow$}}
      & \textbf{\textit{AUC (mean $\pm$\text{ std})}} & \textbf{\textit{$\Delta_{\text{AUC}}\uparrow$}}  & \textbf{\textit{GAUC (mean $\pm$\text{ std})}} & \textbf{\textit{$\Delta_{\text{GAUC}}\uparrow$}} \\
    \midrule
    SASRec & 0.7776 $\pm$ 0.00292 & -
            & 0.6491 $\pm$ 0.00206 & - & 0.6048 $\pm$ 0.00084 & - \\
    DIN     & 0.7806 $\pm$ 0.00118 & +0.30\% 
            & 0.6508 $\pm$ 0.00094 & +0.17\% 
            & 0.6058 $\pm$ 0.00064 & +0.10\%\\
    TA      & 0.7798 $\pm$ 0.00129 & +0.22\% 
            & 0.6538 $\pm$ 0.00046 & +0.47\% 
            & 0.6080 $\pm$ 0.00074 & +0.32\%\\  
    $\text{BFS}_\text{MF}$ & 0.7823 $\pm$ 0.00050 & +0.47\% 
            & 0.6579 $\pm$ 0.00083 & +0.88\%
            & 0.6109 $\pm$ 0.00124 & +0.61\%\\  
    SIMTIER & 0.8090 $\pm$ 0.00233 & +3.14\% 
            & 0.6629 $\pm$ 0.00068 & +1.38\% 
            & 0.6135 $\pm$ 0.00099 & +0.87\%\\  
    MAKE  & 0.8145 $\pm$ 0.00264 & +3.69\% 
             & 0.6623 $\pm$ 0.00075 & +1.32\%
             & 0.6154 $\pm$ 0.00047 & +1.06\%\\  
    DTA    & \underline{0.8214 $\pm$ 0.00184} & \underline{+4.38\%} 
            & \underline{0.6645 $\pm$ 0.00035} & \underline{+1.54\%} 
            & \underline{0.6158 $\pm$ 0.00043} & \underline{+1.10\%} \\
    DMF    & \textbf{0.8251 $\pm$ 0.00105} & \textbf{+4.75\%} 
            & \textbf{0.6663 $\pm$ 0.00049} & \textbf{+1.72\%} 
            & \textbf{0.6177 $\pm$ 0.00060} & \textbf{+1.29\%} \\
    \bottomrule
  \end{tabular}
\end{table*}

\subsection{Datasets}\label{sec:Datasets}

To verify the effectiveness of DMF, we conduct experiments on both public and real-world industrial datasets. For the public dataset, we choose the Amazon\footnote{http://jmcauley.ucsd.edu/data/amazon/} dataset, which has been widely used in CTR prediction~\cite{amazon2019,din2018,distri2025}. For the industrial dataset, we use real-world data collected from the Lazada\footnote{http://www.lazada.com/} recommendation system for experiments.

The Amazon dataset consists of product reviews and metadata from Amazon. We use the Electronics subset to evaluate DMF. As in~\cite{amazon2019}, we treat product reviews as user interaction sequences and organize them chronologically for each user. For a user with a sequence of $T$ interactions, we use the $T$-th interaction as the positive instance and randomly sample another interaction as the negative instance. We also apply 5-core filtering to both users and items.

The industrial dataset was collected from user interaction logs in the recommendation scenario on the international e-commerce platform Lazada. We focus on the subset corresponding to Thailand. The industrial dataset is partitioned temporally, with the instances of past 19 days as the training set and the instances of the 20th day as the test set. The recent 100 interactions are selected as the user interaction sequence.

Basic statistics of these datasets are given in Table~\ref{tab:datasets}.

\begin{table}[tbp]
  \caption{Statistics of datasets.}
  \label{tab:datasets}
  \centering
  \begin{tabular}{cccc}
    \toprule
    Dataset & \texttt{\#}Users &  \texttt{\#}Items & \texttt{\#}Samples\\
    \midrule
    Amazon & 192k & 63k & 1.7M \\
    Industry & 8.7M & 20.9M & 469M \\
    \bottomrule
  \end{tabular}
\end{table}

\subsection{Evaluation Metrics}

For offline experiments, we follow previous work~\cite{din2018,twin2023} to adopt the widely used Area Under Curve (AUC) and Group Area Under Curve (GAUC) for evaluation. GAUC is a variant of AUC, which measures the goodness of intra-user order by averaging AUC over users and is shown to be more relevant to online performance. GAUC is calculated as follows:
\begin{equation}
    \mathrm{GAUC} = \frac{\sum_{u=1}^{U} c_u \cdot \mathrm{AUC}_u}{\sum_{u=1}^{U} c_u},
\end{equation}
where $U$ denotes the number of users, and $c_u$ and $\mathrm{AUC}_u$ denote the number of impressions and the AUC score for the $u$-th user, respectively.

For online experiments, we use post-view Click-Through and ConVersion Rate (CTCVR) and Gross Merchandise Volume (GMV) as online metrics.

\subsection{Baselines}
We compare DMF with the following representative and state-of-the-art user interest modeling approaches for CTR prediction. 

\begin{itemize}
\item \textbf{SASRec}. A target-agnostic user interest modeling approach that adopts the unidirectional self-attention to capture sequential user interests~\cite{sasrec2018}. 
\item \textbf{DIN}. A target-aware user interest modeling approach that uses activation units to model sequential user interests~\cite{din2018}.
\item \textbf{TA}. A target-aware user interest modeling approach derived from standard multi-head attention~\cite{attention2017}, which was originally introduced as MHTA in Section~\ref{sec:ta} and is henceforth abbreviated as TA for brevity.
\item \textbf{$\text{BFS}_\text{MF}$}. An extension of the target attention in TWIN~\cite{twin2023}. TWIN builds a novel attention mechanism via Behavior Feature Splitting (BFS) to efficiently integrate user-item cross features into TA. We adopt the same strategy to incorporate multimodal similarity scores.
\item \textbf{SIMTIER}. A similarity histogram-based approach that converts high-dimensional multimodal representations into a compact similarity histogram. The resulting histogram is treated as a low-dimensional vector and fed into a downstream MLP alongside other embeddings~\cite{make2024}.
\item \textbf{MAKE}. A three-stage framework that employs secondary pre-training to enable convergence of multimodal-related parameters, followed by knowledge transfer to downstream tasks~\cite{make2024}.
\end{itemize}
In summary, SASRec, DIN, and TA are ID-based approaches that rely solely on sparse ID-based features to capture behavioral patterns. In contrast, $\text{BFS}_\text{MF}$, SIMTIER, and MAKE are multimodal approaches that integrate multimodal representations to enhance the effectiveness of user interest modeling.

\begin{table*}[tbp]
  \caption{Ablation Study on Side Information Fusion Strategies. $\Delta_{\text{AUC}}$ is calculated to indicate the average performance boost compared with the baseline (DTA) over the industrial dataset.}
  \label{tab:ablation_study}
  \centering
  \renewcommand{\arraystretch}{1.1}
  \begin{tabular}{ccccc}
    \toprule
    \textbf{Model} & \textbf{AUC (mean $\pm$\text{ std})} & \textbf{$\Delta_{\text{AUC}}\uparrow$}  & \textbf{GAUC (mean $\pm$\text{ std})} & \textbf{$\Delta_{\text{GAUC}}\uparrow$}\\
    \midrule
    TA$_\text{early}$ & 0.6644 $\pm$ 0.00212 & -0.01\%
            & 0.6159 $\pm$ 0.00091 & +0.01\%\\
    TA$_\text{late}$ & 0.6615 $\pm$ 0.00101 & -0.30\%
            & 0.6145 $\pm$ 0.00052 & -0.13\%\\
    DTA$_\text{non-invasived}$ & 0.6624 $\pm$ 0.00079 & -0.21\%
            & 0.6129 $\pm$ 0.00063 & -0.29\%\\
    DTA & 0.6645 $\pm$ 0.00035 & - 
            & 0.6158 $\pm$ 0.00043 & -\\
    \bottomrule
  \end{tabular}
\end{table*}

\subsection{Implementation Details}

In all experiments, models are trained using the Adam optimizer. ID feature embeddings are set to 32 dimensions. For the Amazon dataset, the attention module uses 4 heads and a hidden size of 32, followed by a fully connected network with dimensions 80 × 40 × 1. The batch size is 128. In the industrial dataset, 128-dimensional visual and textual embeddings are fused into a single 128-dimensional vector for multimodal similarity computation. Although this leads to a slight drop in CTR performance, it significantly reduces storage and computational costs. Here, the attention network has 4 heads and a hidden size of 128, with an MLP of 128 × 64 × 1. The batch size is 1024, and the number of buckets $M$ is 35. The secondary pre-training stage of MAKE is conducted for 10 epochs.

\subsection{Performance Comparison}

\subsubsection{Overall Performance(\textbf{RQ1})}

The empirical results in Table~\ref{tab:all_performance} highlight key insights into user interest modeling with ID-only and multimodal features. Target-agnostic models such as SASRec underperform due to their static user representations that ignore target item semantics. In contrast, target-aware models like DIN and TA dynamically reweight historical interactions based on relevance to the target item, achieving consistent gains. 

Incorporating multimodal signals brings further improvements. Methods including $\text{BFS}_\text{MF}$, SIMTIER, and MAKE outperform ID-only baselines. These approaches mitigate modality misalignment through similarity encoding or multi-stage pre-training but process ID and multimodal features separately, limiting fine-grained interaction between content semantics and behavioral signals.

DMF achieves the highest performance among all models, demonstrating the effectiveness of its design. DTA, which fuses ID and multimodal signals through decoupled target-aware attention, already outperforms existing methods, highlighting the value of behavior-level modality interaction. The further gain of DMF over DTA confirms that integrating semantic generalization via histogram modeling with fine-grained behavioral modeling strengthens user representation learning.

\subsubsection{Ablation Study(\textbf{RQ2})}

We conduct an ablation study to analyze the impact of different side information fusion strategies on both model effectiveness, as shown in Table~\ref{tab:ablation_study}. \text{TA$_\text{early}$} represents early fusion by concatenating ID-based embeddings and multimodal similarity scores before computing $\mathbf{K}$ and $\mathbf{V}$. It achieves performance close to our full model, demonstrating the benefit of fine-grained interaction modeling. However, as discussed in Section~\ref{sec:dta}, this approach incurs prohibitive inference costs in practice due to its target-aware computation graph.

In contrast, \text{TA$_\text{late}$} adopts late fusion, where multimodal features are combined only at the final prediction layer. This strategy is computationally efficient but underperforms DTA by $-0.30\%$ in AUC and $-0.13\%$ in GAUC, confirming that delaying fusion limits the capacity of the model to capture fine-grained interactions between content semantics and behavioral signals.

To validate the necessity of our decoupled design, we introduce \text{DTA$_\text{non-invasived}$}, a variant~\cite{nova2011} that injects multimodal similarity signals only into the key ($\mathbf{K}$) path while leaving the value ($\mathbf{V}$) path unchanged. This configuration degrades performance by $-0.21\%$ in AUC and $-0.29\%$ in GAUC, indicating that enriching both $\mathbf{K}$ and $\mathbf{V}$ with content semantic information is essential for effective user interest modeling.

Among all variants, \text{DTA} attains the best AUC and competitive GAUC, while maintaining significantly lower inference latency. This balance of accuracy and efficiency makes DTA particularly suitable for real-world deployment. To evaluate inference efficiency under real-world serving conditions, we perform stress tests on a single NVIDIA A10 GPU with a hidden dimension of 128, user sequence length of 400, and 1200 candidate target items per request, which is representative of industrial ranking workloads. In this setting, the decoupled fusion strategy achieves a 200\% higher throughput (measured in queries per second) compared to early fusion, highlighting its strong suitability for high-throughput, latency-sensitive recommendation systems.

\begin{figure}[tbp]
  \centering
  \begin{minipage}[t]{0.49\columnwidth}
    \centering
    \includegraphics[width=\linewidth]{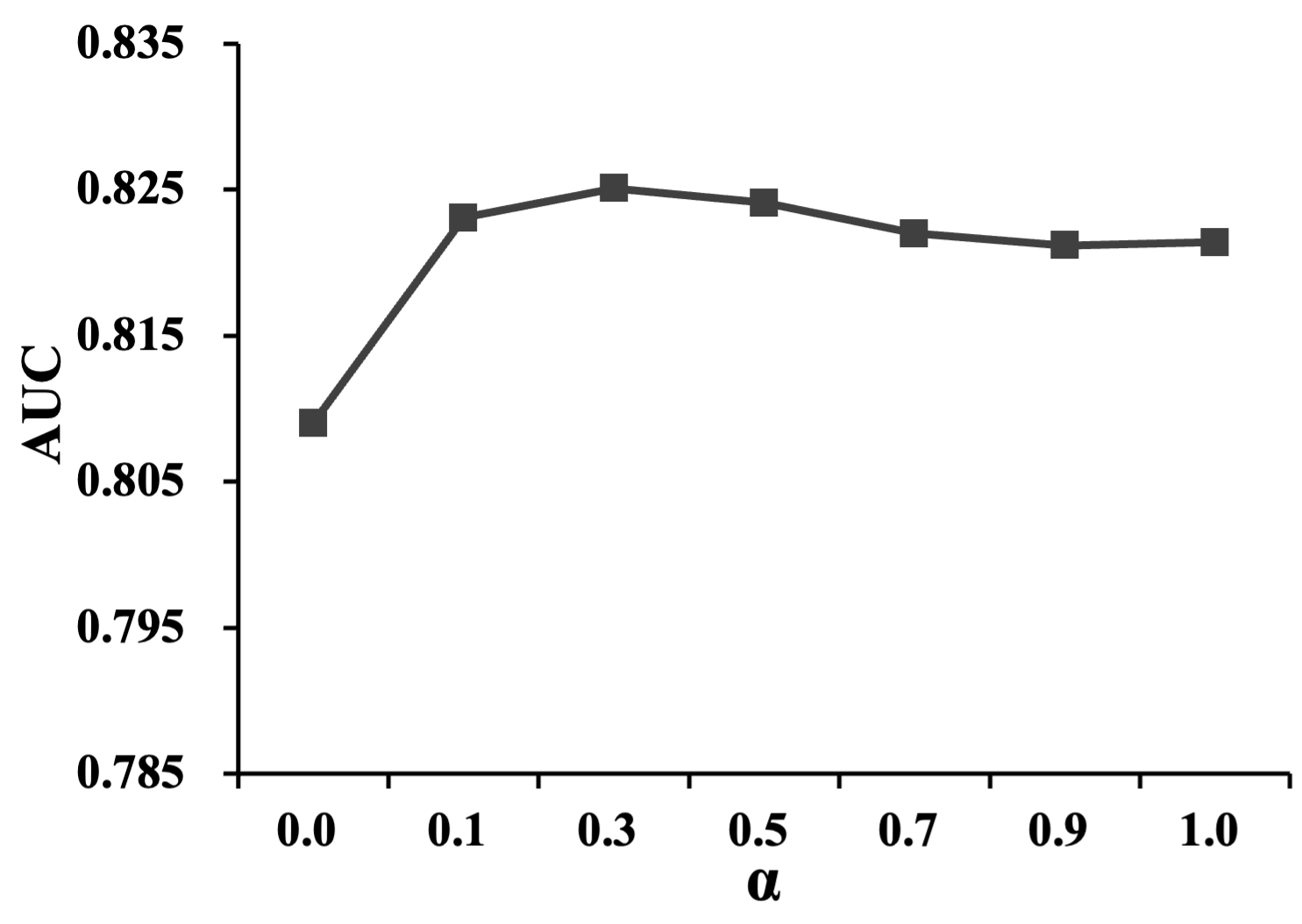}
    \vspace{1mm}
    {\footnotesize (a) Amazon}
    \label{fig:amazon-hy}
  \end{minipage}%
  \hfill
  \begin{minipage}[t]{0.49\columnwidth}
    \centering
    \includegraphics[width=\linewidth]{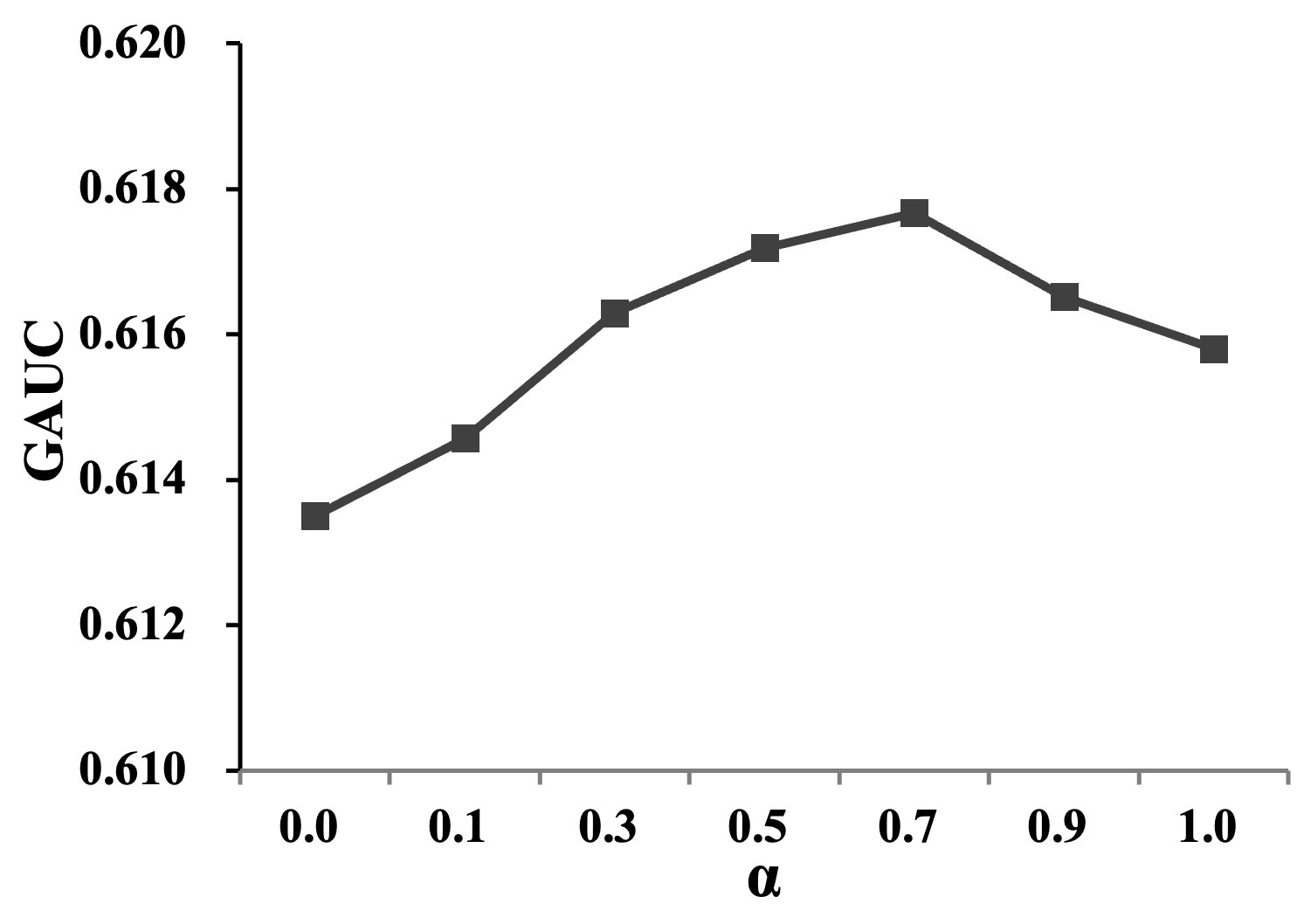}
    \vspace{1mm}
    {\footnotesize (b) Industry}
    \label{fig:industry-hy}
  \end{minipage}
  \caption{Performance with varying representation aggregating hyperparameter $\alpha$. When $\alpha=0$, only the modality-centric modeling strategy is employed, and when $\alpha=1$, only the modality-enriched modeling strategy is utilized.}
  \label{fig:emethod_hy}
\end{figure}

\subsubsection{Hyper-parameters Study(\textbf{RQ3})}\label{sec:parameters}

The representation aggregating hyperparameter $\alpha$ controls the trade-off between the modality-enriched representation $\text{R}_{me}$ and the modality-centric representation $\text{R}_{mc}$, effectively adjusting the relative contribution of each modeling strategy to the final user interest representation $\text{R}_u$. Fig.~\ref{fig:emethod_hy} illustrates the sensitivity of DMF to $\alpha$ on Amazon and industrial datasets. We find that the worst performance consistently occurs when $\alpha=0$, as in this setting, DMF relies solely on $\text{R}_{mc}$, completely disregarding the fine-grained behavioral signals captured by $\text{R}_{me}$, indicating that content-driven generalization alone, while robust, is insufficient for optimal personalization. As $\alpha$ increases, the influence of $\text{R}_{me}$ grows, leading to consistent improvements in prediction performance. The Amazon dataset shows the peak performance at $\alpha=0.3$, and the industrial dataset shows the peak performance at $\alpha=0.7$, suggesting that the optimal balance favors $\text{R}_{me}$ but still requires substantial contribution from $\text{R}_{mc}$ to maintain semantic coherence and generalization capability. At $\alpha = 1$, DMF depends entirely on $\text{R}_{me}$, which fuses item identity signals with multimodal context. However, ID-based embeddings embedded within this modality-enriched modeling strategy can be biased or stale, especially under interest shift or adversarial behavior, leading to suboptimal performance.

\begin{figure}[tbp]
  \centering
  \begin{minipage}[t]{0.49\columnwidth}
    \centering
    \includegraphics[width=\linewidth]{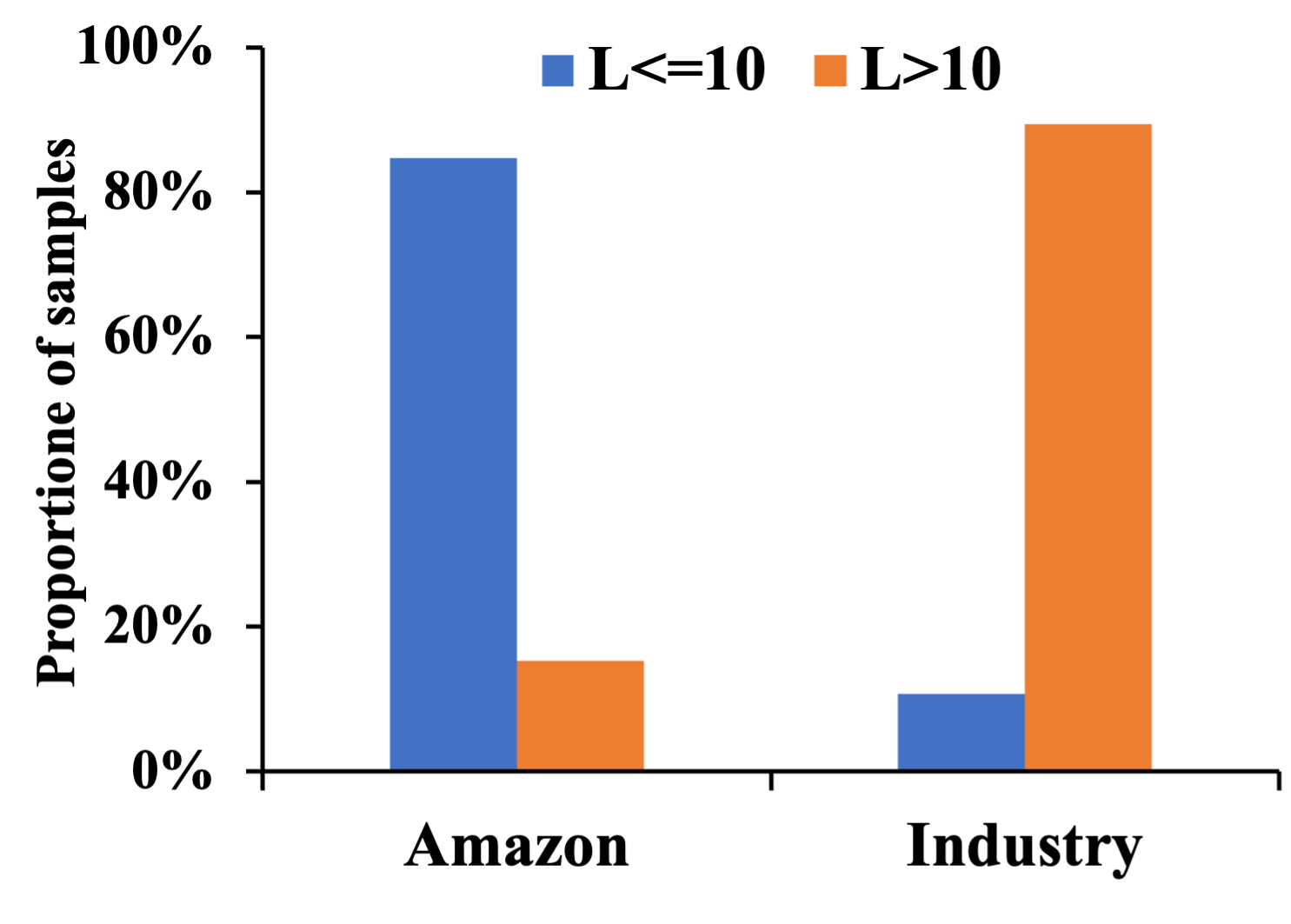}
    \vspace{1mm}
    {\footnotesize (a)}
    \label{fig:sample-user}
  \end{minipage}%
  \hfill
  \begin{minipage}[t]{0.49\columnwidth}
    \centering
    \includegraphics[width=\linewidth]{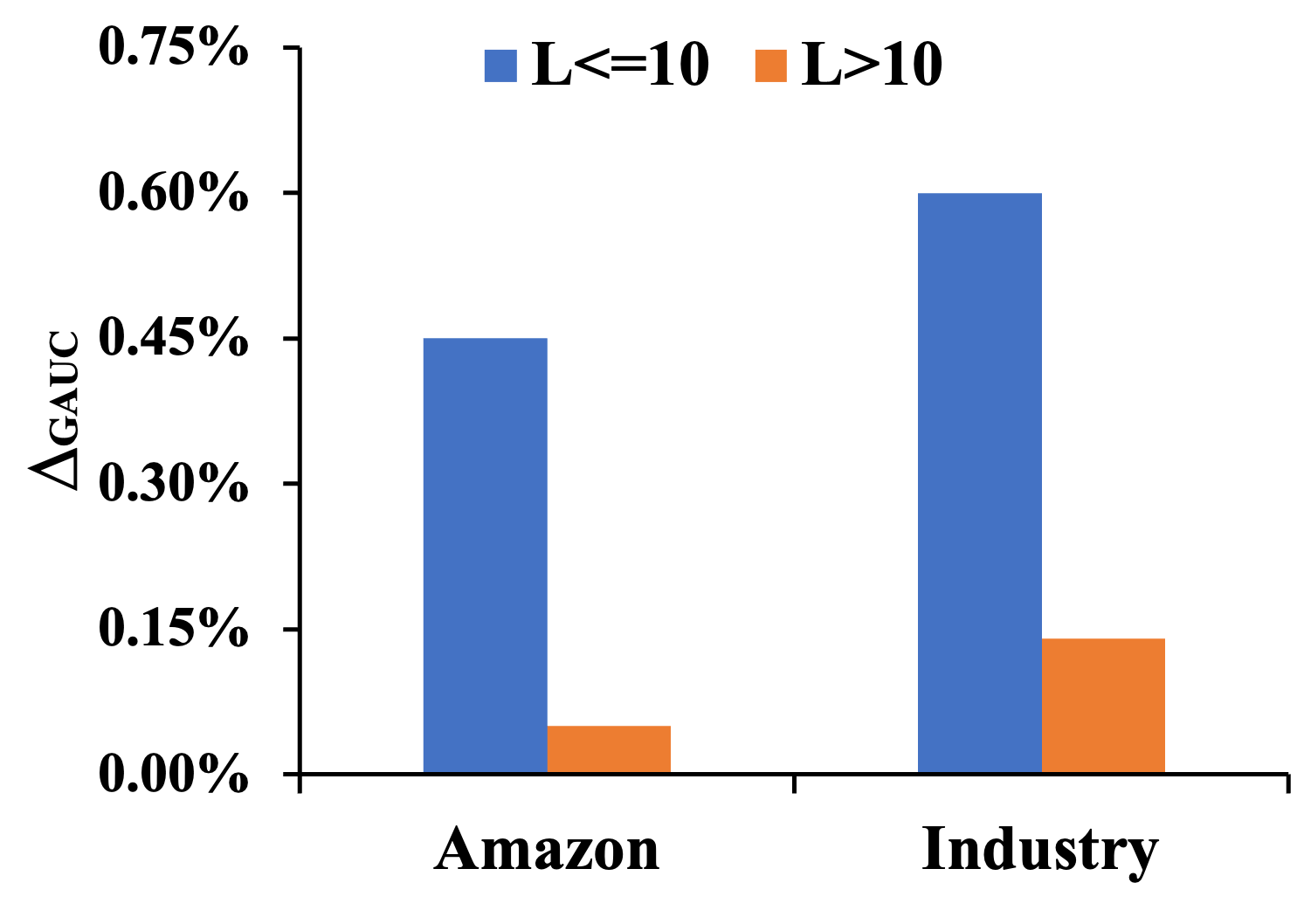}
    \vspace{1mm}
    {\footnotesize (b)}
    \label{fig:gauc-user}
  \end{minipage}
  \caption{Relationship between user interaction sequence length and model performance. (a) Proportion of samples for users with short ($L\leq10$) and long ($L>10$) interaction sequences on Amazon and Industry datasets.(b) GAUC improvements achieved by DMF over DTA for users with short ($L\leq10$) and long ($L>10$) interaction sequences on Amazon and Industry datasets.}
  \label{fig:analysis-user}
\end{figure}

\begin{figure*}[tbp]
\centering
\includegraphics[width=1.0\linewidth]{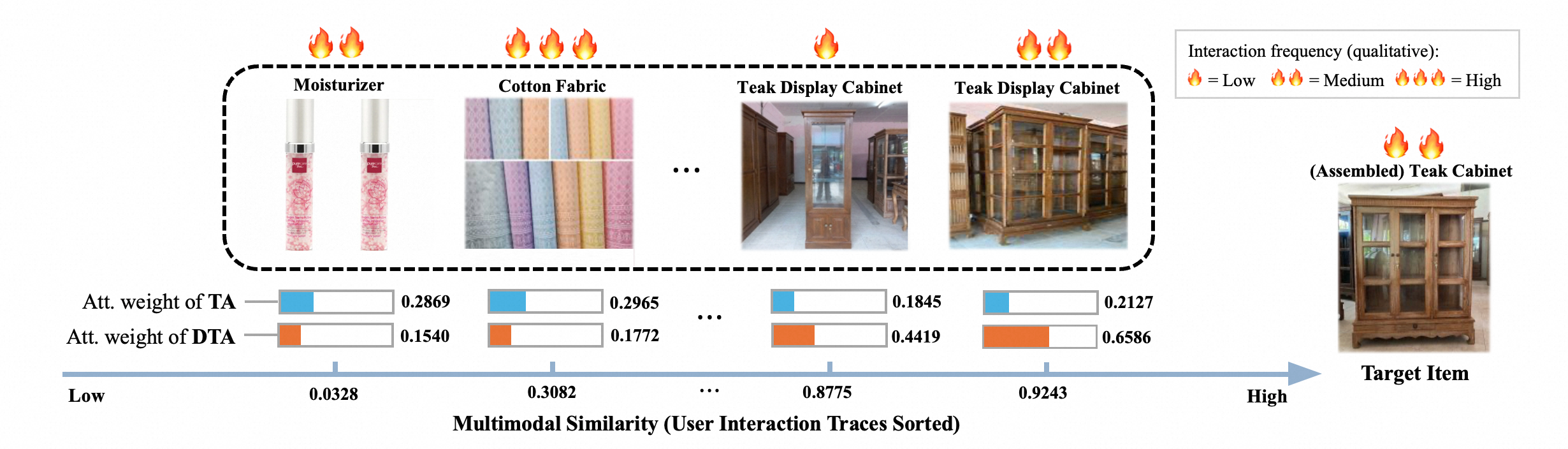}
\caption{Illustration of the attention weights in TA and DTA. Interactions with high relevance to the target item get high attention weights.}
\label{fig:case}
\end{figure*}

\subsubsection{Analysis of Representation Aggregation Dynamics(\textbf{RQ4})}\label{sec:aggregation_dynamics}

An interesting question is why the optimal value of $\alpha$ varies significantly between Amazon and industrial datasets. Fig.~\ref{fig:analysis-user} reveals that the modality-centric representation $\mathbf{r}_{\mathrm{mc}}$ delivers substantially higher GAUC gains for users with short interaction sequences ($L \leq 10$) than for those with long sequences ($L > 10$), consistently across both datasets. This effect explains the divergent optimal $\alpha$ values: the Amazon dataset, where 84.75\% of users are low-activity ($L \leq 10$), achieves peak performance at $\alpha = 0.3$, while the industrial dataset, dominated by high-activity users (89.35\% with $L > 10$), favors $\alpha = 0.7$ to prioritize modality-enriched modeling.

This pattern suggests that semantic generalization is more effective under conditions of action sparsity. For low-activity users, limited historical interactions constrain the reliability of behavior-driven personalization; here, modality-centric features provide a robust semantic prior that mitigates data scarcity. In contrast, high-activity users generate abundant behavioral signals, enabling modality-enriched representations to precisely align recommendations with fine-grained preferences by fusing ID-based collaboration with multimodal context. Thus, the optimal fusion weight $\alpha$ is not an arbitrary hyperparameter but a reflection of the user population’s activity profile.

From a system design perspective, this insight advocates for activity-aware fusion strategies in industrial recommenders. Rather than using a global $\alpha$, platforms should adapt the aggregation weight based on user activity level, that is, assigning a higher weight to $\mathbf{r}_{\mathrm{mc}}$ for new users or low-activity user groups and to $\mathbf{r}_{\mathrm{me}}$ for high-activity user groups. Such an approach would dynamically balance robustness and personalization according to the informational sufficiency of each user interaction history, transforming a static model component into an intelligent, context-responsive mechanism that serves diverse user groups effectively.

\subsubsection{Case Study(\textbf{RQ5})}

We conduct case study to reveal the inner structure of DTA on the industrial dataset. Since attention weights computed via softmax normalization are not comparable across models, we use sigmoid normalization to compute attention weights for both TA and DTA. Fig.~\ref{fig:case} visualizes the attention weights assigned by TA and DTA to user interaction traces with respect to a target item.

As shown in Fig.~\ref{fig:case}, TA assigns the highest weight $0.2965$ to Cotton Fabric, which is a highly popular item with frequent interactions indicated by three flame icons. However, this item exhibits low semantic relevance to the target item, which is an assembled teak cabinet. In contrast, Teak Display Cabinet, a less frequently interacted item indicated by two flame icons, shares high multimodal similarity $0.9243$ with the target item, yet receives only $0.2127$ attention weight under TA, reflecting its inability to capture content-level relevance.

DTA integrates multimodal cosine similarity as a target-aware signal to modulate attention computation. DTA correctly identifies that Teak Display Cabinet is semantically aligned with the target item and assigns it the highest attention weight ($0.6586$), significantly surpassing all other items in the sequence. This demonstrates the capability of DTA to transcend popularity bias and prioritize behaviorally rare but semantically relevant interactions. 

\subsubsection{Online A/B Testing (\textbf{RQ6})}

We apply the DMF framework to the ranking model in the product recommendation system of the international e-commerce platform Lazada, ventures of Thailand. Each target item in the ranking system will be ranked based on a rank score, which is estimated based on the CTR and CVR (Conversion Rate).

The online base model used TA to extract user interests from user interaction sequences and used SIMTIER to integrate the pre-trained multimodal representations, including the representations for both visual and textual modalities. Compared with the base model, we substitute the TA and SIMTIER modules with DMF to model behavioral patterns and multimodal signals together. After conducting an online A/B test for 12 days, DMF demonstrated promising results, achieving relative improvements of 5.30\% in CTCVR and 7.43\% in GMV. The deployment introduces moderate additional computation per request; however, this is efficiently absorbed through horizontal scaling without increasing end-to-end latency. 
The associated increase in resource cost is marginal compared to the magnitude of business gains, demonstrating the favorable return on investment and deployability at scale of DMF.

\begin{table}[tbp]
  \caption{Online A/B testing results. The improvements on GMV and CTCVR are reported. }
  \label{tab:abtest}
  \centering
  \begin{tabular}{ccc}
    \toprule
    \textbf{Metric} & \textbf{GMV} & \textbf{CTCVR}\\
    \midrule
    Lift rate & +7.43\% & +5.30\% \\
    \bottomrule
  \end{tabular}
\end{table}

\section{Conclusion}\label{sec:conclusion}

We proposed Decoupled Multimodal Fusion (DMF), a hybrid framework for user interest modeling in CTR prediction. By introducing Decoupled Target Attention (DTA), DMF enables fine-grained interaction between ID-based and multimodal representations while preserving inference efficiency through computational decoupling. The Complementary Modality Modeling (CMM) module fuses modality-centric and modality-enriched representations for more comprehensive interest modeling. Experiments show that DMF outperforms state-of-the-art methods on both public and industrial datasets. Deployed on the product recommendation system of the international e-commerce platform Lazada, DMF achieves significant improvements of +5.30\% in CTCVR and +7.43\% in GMV, demonstrating its effectiveness and practical value in real-world applications.

\bibliographystyle{IEEEtran}
\bibliography{icde2026cutrefs}

\end{document}